\documentclass[twocolumn,prl]{revtex4}
\usepackage{natbib}
\usepackage{amsmath}
\usepackage{graphicx}
\renewcommand{\vec}[1]{\ensuremath{\mathchoice%
       {\mbox{\boldmath$\displaystyle#1$}}
       {\mbox{\boldmath$\textstyle#1$}}
       {\mbox{\boldmath$\scriptstyle#1$}}
       {\mbox{\boldmath$\scriptscriptstyle#1$}}}}

\begin{document}
\title{The Lagrangian frequency spectrum as a diagnostic for magnetohydrodynamic 
turbulence dynamics}
\author{Angela Busse} 
\email{angela.busse@ipp.mpg.de}
\thanks{now at University of Southampton, United Kingdom}
\author{Wolf-Christian M\"{u}ller}
\email{wolf.mueller@ipp.mpg.de}
\affiliation{Max-Planck-Institut f\"{u}r Plasmaphysik, Boltzmannstr.\ 2,
85748 Garching bei M\"unchen, Germany}
\author{Grigol Gogoberidze}
\email{gogober@geo.net.ge}
\affiliation{Georgian National Astrophysical Observatory, 2a Kazbegi Avenue, 0160 Tbilisi Georgia}
\affiliation{Centre for Plasma Astrophysics, University of Leuven}
\date{\today}
\begin{abstract}
For the phenomenological description of magnetohydrodynamic turbulence 
competing models exist, e.g. Boldyrev [Phys.Rev.Lett. \textbf{96}, 115002, 2006] and Gogoberidze [Phys.Plas. \textbf{14}, 022304, 2007], which predict the same Eulerian inertial-range
scaling of the turbulent energy spectrum although they employ fundamentally
different basic interaction mechanisms. 
{A relation is found that links} the Lagrangian frequency spectrum  
{with} the autocorrelation timescale of the turbulent fluctuations, 
$\tau_\mathrm{ac}$, and the associated cascade timescale,
$\tau_{\mathrm{cas}}$. 
Thus, the Lagrangian energy spectrum can serve to identify weak
($\tau_\mathrm{ac}\ll\tau_{\mathrm{cas}}$) and strong
($\tau_\mathrm{ac}\sim\tau_{\mathrm{cas}}$) interaction mechanisms 
providing insight into the turbulent energy cascade.
The new approach is illustrated by results from direct 
numerical simulations of two- and three-dimensional incompressible 
MHD turbulence. 
\end{abstract}
\maketitle
Magnetohydrodynamic (MHD) turbulence has been the subject of intense
research during the last decades since turbulent low-frequency,
long-wavelength fluctuations conveniently described in the
incompressible MHD framework are present in many astrophysical systems
(see for example \cite{biskamp:book3} and references therein). 
As of now, the theoretical description of the universal statistical
properties of turbulent flows relies mainly on phenomenological models
like Kolmogorov's K41 picture \cite{frisch:book} of hydrodynamic
turbulence.  In the incompressible MHD case the phenomenological
description is complicated by the presence of shear Alfv\'en wave
modes and their associated timescale, the Alfv\'en time $\tau_\text{A}$.
Especially for turbulence in the presence of a strong  mean
magnetic field there exist competing phenomenological models of the
inertial-range energy cascade 
\cite{iroshnikov:ikmodel,kraichnan:ikmodel,
goldreich_sridhar:gs2,boldyrev:boldymodelII,gogoberidze:gogomodel}.  
Although based on physically different
mechanisms the most recent models by Boldyrev \cite{boldyrev:boldymodelII} 
and Gogoberidze \cite{gogoberidze:gogomodel} 
predict identical diagnostic signatures, e.g., the
same inertial-range scaling of the energy spectrum, and are thus
hardly distinguishable by conventional measurements in the Eulerian
frame of reference. 
Complementing Eulerian
diagnostics, the Lagrangian frequency spectrum gives insight into the
timescales associated with the turbulent energy cascade and the underlying 
nonlinear interactions of turbulent fluctuations.

This Letter reports a fundamental relation between the Lagrangian
frequency spectrum and the characteristic timescales of turbulence, the
autocorrelation time $\tau_\mathrm{ac}$ and the cascade time
$\tau_\mathrm{cas}$. 
Here, the autocorrelation time $\tau_{\mathrm{ac}}=\tau_{\mathrm{ac}}(\ell)$ 
characterizes the dominant nonlinear interaction process between turbulent fluctuations 
with $\ell$ being the spatial scale under consideration.
On the
cascade time scale $\tau_{\mathrm{cas}}$ the fluctuations at a fixed spatial scale loose their coherence and decay into smaller
turbulent fluctuations.
The relation presented in the following
allows to investigate a fundamental
aspect of turbulent dynamics, i.e. to distinguish whether nonlinear
interactions are weak, $\tau_\mathrm{ac}\ll\tau_\mathrm{cas}$, or
strong $\tau_\mathrm{ac}\sim\tau_\mathrm{cas}$.  
The relation is supported
by high-Reynolds-number direct numerical simulations 
of
two- and three-dimensional MHD turbulence.  
In the following, pertinent turbulence phenomenologies
are summarized 
focusing on their characteristic
timescales.

In the K41 picture of Navier-Stokes turbulence the
auto-correlation timescale $\tau_\mathrm{ac}$ is determined
dimensionally by the nonlinear turnover time
$\tau_\mathrm{NL}=\ell/v_\ell$ where $v_\ell$ is a
velocity fluctuation (eddy)  at scale $\ell$.  A single nonlinear interaction
between eddies reduces the coherence of an involved fluctuation so
significantly that it ceases to exist at scale $\ell$ having generated
fluctuations at slightly smaller scales.
The required time for the associated decorrelation in the K41
picture is $\tau_\mathrm{cas}\sim\tau_{\mathrm{NL}}\sim\tau_\mathrm{ac}$, thus
the nonlinear interaction is {strong}.
In incompressible two-dimensional MHD turbulence the
Iroshnikov-Kraichnan (IK) phenomenology
\cite{iroshnikov:ikmodel,kraichnan:ikmodel} seems to apply, see e.g.
\cite{biskamp_welter:2dmhddecay}.  There, colliding counter-propagating
Alfv\'enic fluctuations lead to a wave-based nonlinear decorrelation
of turbulent structures. The fundamental interaction timescale is the
Alfv\'en-time $\tau_\mathrm{A}=\ell/b_0$ where 
the assumption of constant mass density  
yields the  Alfv\'{e}n speed 
as the value of a properly normalized external magnetic guide field $b_0$ or, 
if no such field is present,
the slowly varying large-scale magnetic fluctuations $b_{\mathrm{rms}}$.
As colliding Alfv\'en wave packets experience only a
small deformation in a single nonlinear interaction, 
many consecutive interactions
are required for a cascade step, i.e.
$\tau_\mathrm{A}\ll\tau_\mathrm{NL}^2/\tau_\mathrm{A}\sim\tau_\mathrm{cas}$
since generally $\tau_\mathrm{A}\ll\tau_\mathrm{NL}$, and the nonlinear
interaction is {weak}. 
In incompressible three-dimensional MHD turbulence with {weak to moderate}
mean magnetic field the Goldreich-Sridhar (GS)
\cite{goldreich_sridhar:gs2} phenomenology has become widely accepted.
It
enhances the IK picture by explicitly taking into account dynamical
anisotropy with regard to the direction of the local magnetic field. In
addition, the hypothesis of a critical balance of turnover and Alfv\'en
time is made,
i.e. $\tau_\mathrm{ac}\sim\tau_\mathrm{A}\sim\tau_\mathrm{NL}\sim\tau_\mathrm{cas}$
resulting in strong nonlinear interaction.

However, the inertial-range scaling of the Eulerian energy spectrum
of MHD turbulence in a strong mean magnetic field
does not agree with the GS phenomenology \cite{maron_goldreich:anisomhd,mueller_biskamp_grappin:anisomhd, mueller_grappin:endyn,mason_cattaneo_boldyrev:specmeas512,perez_boldyrev:crosshelmhd}. 
A 
proposed
alternative is the dynamic-alignment model (DA)
\cite{boldyrev:boldymodelII} which extends the GS model by a
scale-dependent polarization of the interacting wave packets but still leads to 
strong nonlinear interaction. By a  different approach, an anisotropic variant of the
IK-picture (AIK) \cite{gogoberidze:gogomodel} 
yields weak
interaction dynamics. {It proposes nonlocal decorrelation effects on inertial-range scales by large-scale 
fluctuations.} 
Both phenomenologies lead to identical scaling 
results with regard to Eulerian
two-point statistics although their underlying physical assumptions are fundamentally  different.

The inertial-range scaling of the Eulerian energy spectrum $E(k)$ 
serves as a standard diagnostic for turbulence investigations, but  
is in fact not unique:
the K41 and GS models yield $E(k)\sim k^{-5/3}$. In the GS picture 
the wavenumber $k$ is defined in a direction perpendicular to the magnetic field, 
$k\rightarrow k_\perp$. 
The DA, AIK and IK models, in contrast, all 
predict $E(k)\sim k^{-3/2}$ with $k\rightarrow k_\perp$ for DA and AIK.
This 
has motivated the development of a new diagnostic approach that
allows to probe the relation between $\tau_{\mathrm{ac}}$ and $\tau_{\mathrm{cas}}$
as presented in the following.

The Lagrangian  two-point two-time velocity correlation  \cite{kaneda:eullag} is defined as  
\begin{equation}
R_L=\langle \vec{V}(\vec{X}_{0}+\vec{r},\tau+t_{0})\cdot \vec{V}(\vec{X}_{0},t_{0})\rangle,\label{lagracorr}
\end{equation} 
where $\vec{V}(\vec{X}_{0}+\vec{r},t_{0}+\tau)$  is the velocity measured at time $t_0+\tau$
of a fluid element {that was at position $\vec{X}_0+\vec{r}$ at time $t_0$. The Lagrangian
variable $\vec{X}=\vec{X}(\vec{X}_0,t)$ is the time-dependent position
along a fluid particle's trajectory. The Lagrangian velocity
$\vec{V}(\vec{X}_0,t)$ is connected to the Eulerian velocity field
by $\vec{V}(\vec{X}_0,t)=\vec{v}(\vec{x}=\vec{X}(\vec{X}_0,t),t)$.}
While $R_\mathrm{L}$ is generally a correlation function involving two fluid particles 
at positions  $\vec{X}^{(1)}(\vec{X}_{0}+\vec{r},\tau+t_0)$ and $\vec{X}^{(2)}(\vec{X}_{0},t_0)$ 
it reduces to a two-point velocity correlation 
along a single trajectory if $\vec{r}=0$.
The corresponding Eulerian correlation function is
\begin{equation}
R_{\mathrm{E}}=
\langle \vec{v}(\vec{x}+\vec{r},t+\tau)\cdot \vec{v}(\vec{x},t)\rangle.\label{eulercorr}
\end{equation}
with $\vec{v}(\vec{x},t)$ representing the velocity field at a position $\vec{x}$ and time $t$ while 
$\vec{r}$ and $\tau$ stand for independent translations in space and time, respectively. 

For statistically homogeneous and stationary turbulence these functions depend on 
$\vec{r}$ and $\tau$ only, $R_{\mathrm{L,E}}=R_{\mathrm{L,E}}(\vec{r},\tau)$.
The corresponding two-time spectral functions $Q(\vec{k},\tau)$ are defined as the
Fourier transforms of the correlation functions
 \begin{equation}
R_{\mathrm{L,E}}=\int d^3 \vec{k} \exp(i \vec{k}\cdot \vec{r}) Q_{\mathrm{L,E}}(\vec{k},\tau)\,.
\label{specfunc}
\end{equation}
The two-time spectral functions are related to the three-dimensional energy spectrum
${E}(\vec{k})$ by 
\begin{equation}
Q_{\mathrm{L,E}}(\vec{k},\tau)={E}(\vec{k})G_{\mathrm{L,E}}(\tau/\tau_{\mathrm{ac}}^{\mathrm{L,E}})\,, \label{lesrel}
\end{equation}
where $\tau_{\mathrm{ac}}^{\mathrm{L,E}}$ are the Lagrangian and Eulerian autocorrelation time
scales, and $G_{\mathrm{L,E}}(\tau/\tau_{\mathrm{ac}}^{\mathrm{L,E}})$ are the corresponding
response functions \cite{leslie:book}. 
The general features of models of the response function used in the theoretical
description of turbulence \cite{kraichnan:dia,edwards:generalrandomphase,leslie:book} 
are that $G_{\mathrm{L,E}}$ is a smooth function
with $G_{\mathrm{L,E}}(0)=1$, $G_{\mathrm{L,E}}(x)=0$ for $x\gg 0$ and
$\int_{0}^{\infty}G_{\mathrm{L,E}}(x)dx=1$.

The Lagrangian and Eulerian frequency spectra are defined as \cite{kaneda:eullag}
\begin{equation}
\Phi_\mathrm{L,E}(\omega)=\frac{1}{2\pi}\int d\tau \cos(\omega\tau)R_\mathrm{L,E}(0,\tau)\,.
\label{phispecs}
\end{equation}
{
Putting Eq. (\ref{lesrel}) in Eq. (\ref{specfunc}), and plugging the result in Eq. (\ref{phispecs})} 
yields for the frequency spectrum 
\begin{equation} 
\Phi_\mathrm{L,E}(\omega)=\frac{1}{\pi}\int \mathrm{d}\tau \int \mathrm{d}k E(k) G_\mathrm{L,E}(\tau/\tau_{\mathrm{ac}}^\mathrm{L,E})\cos(\omega \tau)
\end{equation}
with $E^\mathrm{total}=\int \mathrm{d}k E(k)$. {The energy spectrum $E(k)$ and the associated wavenumber has
to be defined in a suitable way for 
the considered geometry e.g. spherical, planar, or cylindrical.}  
Under the assumption of self-similarity of all involved dependent variables a dimensional approximation of this result linking wavenumber and frequency energy spectra in a simple way 
yields
\begin{equation}
\omega\Phi_\mathrm{L,E}(\omega)\sim kE(k)\quad\text{with} \quad\omega\sim 1/\tau^\mathrm{L,E}_\mathrm{ac}(k)\,.\label{mainres}
\end{equation}
Note that relation (\ref{mainres}) is also found in \cite{tennekes:freqspec2} following a different and  more specific 
approach.

The Eulerian correlation time at a fixed position 
is dominated by the sweeping of small-scale fluctuations by the 
largest-scale eddies 
$\tau_\mathrm{ac}^\mathrm{E}\sim (kv_0)^{-1}$. Consequently, the spectral scaling of frequency and wavenumber spectra
should be identical in this case with $\Phi^\mathrm{GS}_\mathrm{E}(\omega)
\sim(\varepsilon v_0)^{2/3}\omega^{-5/3}$  or 
$\Phi_\mathrm{E}^\mathrm{IK}(\omega)\sim \varepsilon^{1/2}v_0\omega^{-3/2}$ where the choice depends on the respective 
Eulerian inertial-range scaling. 
For MHD turbulence, however, the Lagrangian frequency spectrum 
allows to distinguish turbulence phenomenologies based on strong nonlinear interaction 
such as GS and DA
from pictures based on
inherently weak interaction 
like (A)IK. 
Here, the Lagrangian autocorrelation
time is assumed to be characteristic for nonlinear interactions in the energy cascade. In the MHD case,
this holds as long as spectral kinetic and magnetic energy are sufficiently close to equipartition.
A detailed investigation of the influence of the large wavenumber contribution of the Eulerian 
spectrum on the Lagrangian inertial range \cite{kaneda:eullag} shows that the 
scaling of the Lagrangian inertial range is not adulterated by the Eulerian large-scale interval.   

With $\omega\sim 1/\tau^\mathrm{L}_\mathrm{ac}(k)$ and $E(\ell)\sim\omega\Phi_\mathrm{L}(\omega)$,
the classical constant-flux ansatz $E(\ell)/\tau_\mathrm{cas}\sim\varepsilon$
gives  
$\Phi_\mathrm{L}(\omega)\sim \varepsilon \tau_\mathrm{cas}\tau^\mathrm{L}_\mathrm{ac}$.
Strong interaction turbulence is characterized by $\tau_\mathrm{ac}\sim\tau_\mathrm{cas}$ 
resulting in 
\begin{equation}
\Phi_\mathrm{L}(\omega)\sim \varepsilon \omega^{-2}\,.\label{2fspec}
\end{equation}
This scaling is thus expected  for the GS and the DA model and is well-known from Navier-Stokes
turbulence \cite{tennekes:freqspec,yeung_pope_sawford:2048simul}.
{
In contrast, for IK-based phenomenologies with a weak interaction mechanism $\tau_\mathrm{ac}\ll\tau_\mathrm{cas}\sim 
\tau^2_\mathrm{NL}/\tau^\mathrm{L}_\mathrm{ac}$. 
Using relation (\ref{mainres}), $\tau_\mathrm{ac}^\mathrm{L}\sim\omega^{-1}\sim (kb_0)^{-1}$, and $\tau_\mathrm{NL}\sim (k^3E_k)^{-1/2}$ yields $\tau_\mathrm{cas}\sim b_0^2\Phi^{-1}_\mathrm{L}\omega^{-2}$. Thus}
\begin{equation}
\Phi_\mathrm{L}(\omega)\sim\varepsilon^{1/2}b_0\omega^{-3/2}\,.\label{32fspec}
\end{equation} 
Note that this result holds for IK. In the case of AIK 
the quasi-constant large-scale magnetic field  $b_0$ has 
to be replaced by $v_\mathrm{rms}$.
 
Lagrangian frequency spectra are obtained by tracking fluid particles
in direct numerical pseudospectral simulations of MHD turbulence. 
Details of the numerical method can be found in 
\cite{mueller_biskamp_grappin:anisomhd,busse_mueller_homann_grauer:lagradisp}.
In the three-dimensional simulations the number of tracers amounts to $3.2\cdot 10^6$, 
except for the lowest resolution runs where it is lowered to $5\cdot 10^5$.  
In the two-dimensional simulations $2\cdot 10^{6}$ tracers have been tracked.
Important parameters and characteristics of the simulations are listed in Table \ref{param-table}.
The magnetic Prandtl number $\mathsf{Pr_m}$, the ratio of kinematic viscosity $\nu$ to magnetic 
diffusivity $\eta$, is  unity. {Significant deviations from this value would lead to a degradation of
observable inertial-range self-similarity due to numerical resolution constraints and are thus not considered here.} 
\begin{table}
\begin{tabular}{cccccccc}
Re & $u_{\mathrm{rms}}$ & $b_{\mathrm{rms}}$ & $b_{0}$ & $\varepsilon^{\mathrm{K}}$
& $\varepsilon^{\mathrm{M}}$ & $\nu$ & $N_\mathrm{colloc}$\\ \hline
$2150$ & $0.75$ & $0.93$ & $0$ & $6.4\cdot10^{-2}$ & $9.7\cdot10^{-2}$ & $5\cdot 10^{-4}$ & $1024^2$ \\
$6200$ & $0.80$ & $0.98$ & $0$ & $7.2\cdot10^{-2}$ & $1.0\cdot10^{-1}$ & $1.8\cdot 10^{-4}$ & $2048^2$ \\
$18240$ & $0.80$ & $0.98$ & $0$ & $6.8\cdot10^{-2}$ & $8.9\cdot10^{-2}$ & $6\cdot 10^{-5}$  & $4096^2$ \\
$1050$ & $0.44$& $0.59$ & $0$ & $0.11$ & $0.17$ & $1\cdot 10^{-3}$  & $512^3$ \\ 
$3150$ & $0.46$ & $0.64$ & $0$ & $0.12$& $0.17$ & $3.4\cdot 10^{-4}$ & $1024^3$ \\
$1790$ & $0.53$ & $0.62$ & $5$ & $7.4\cdot 10^{-2}$ & $8.5\cdot 10^{-2}$& $8\cdot 10^{-4}$ & $512^2\cdot 256$ \\
$4410$ & $0.55$ & $0.63$ & $5$ & $7.7\cdot 10^{-2}$ & $8.8\cdot 10^{-2}$& $3.3\cdot 10^{-4}$ & $1024^2\cdot 512$ \\
\end{tabular}
\caption{\label{param-table}
Parameters of the numerical simulations. Re: Reynolds number, $u_\mathrm{rms}$; $b_\mathrm{rms}$: RMS value of velocity 
and magnetic field fluctuations;
$b_0$: external mean magnetic field; 
$\varepsilon^{\mathrm{K}}$, $\varepsilon^{\mathrm{M}}$: kinetic and magnetic energy dissipation rates; 
$\nu$ kinematic viscosity; $N_\mathrm{colloc}$: numerical resolution.
}
\end{table}
In the macroscopically isotropic cases, $b_0=0$, both the magnetic and velocity
field are forced by independent Ornstein-Uhlenbeck processes \cite{ornstein_uhlenbeck:process,eswaran_pope:forcing} 
in the wavenumber
shell $k_\mathrm{f}=3$ in order to maintain quasistationary turbulence.
In the anisotropic MHD case, $b_0=5$, large scale Alfv\'{e}nic fluctuations are
excited by the stochastic forcing method which impair 
the frequency scaling range.
Therefore in this case the system is forced by freezing the lowest
wavenumber modes of the velocity and magnetic field of a fully developed
turbulent state, see, e.g.,  \cite{mueller_grappin:endyn}. 
{In the two-dimensional case turbulence is maintained
by keeping the kinetic and magnetic energy in the lowest
wavenumber shells  $1\le |k| < 3=k_\mathrm{f}$ at a constant value.}
Based on the estimate $(L_0/\ell_\mathrm{diss})^{4/3}$ for the width of the inertial-range, the 
Reynolds number is defined as 
$Re= [2\pi/(k_\mathrm{f} \ell_\mathrm{diss})]^{4/3}$ where 
$\ell_\mathrm{diss}\sim (\nu^3/\varepsilon^\mathrm{K})^{1/4}$ is the Kolmogorov dissipation length.
To obtain the Lagrangian frequency spectrum first the autocorrelations of the
velocity and the magnetic field fluctuations along the particle trajectories
are calculated. The spectrum is then computed 
as the cosine transform of the {auto-}correlation functions \cite{yeung_pope:trackalgo}
\begin{equation}
\label{freqspecesdef}
\Phi_{\mathrm{L},i}(\omega)=\frac{1}{2\pi} \int d\tau \left(\langle V_{i}(t+\tau) V_{i}(t)\rangle +
\langle b_{i}(t+\tau) b_{i}(t)\rangle\right)\cos(\omega\tau)\,.
\end{equation}
In the macroscopically isotropic MHD cases the frequency spectrum does 
not depend
on the component $i$ of the velocity or the magnetic field fluctuations.
Here the total frequency spectrum $\Phi(\omega)=\sum_{i=1}^{3}\Phi_{i}(\omega)$ 
is shown.
In the hydrodynamic case (not shown) the spectrum 
scales as $\omega^{-2}$ in the inertial
range in agreement with previous
experimental \cite{mordant_leveque_pinton:experiment} and numerical 
\cite{yeung_pope_sawford:2048simul} results.
\begin{figure}
\includegraphics[width=8cm]{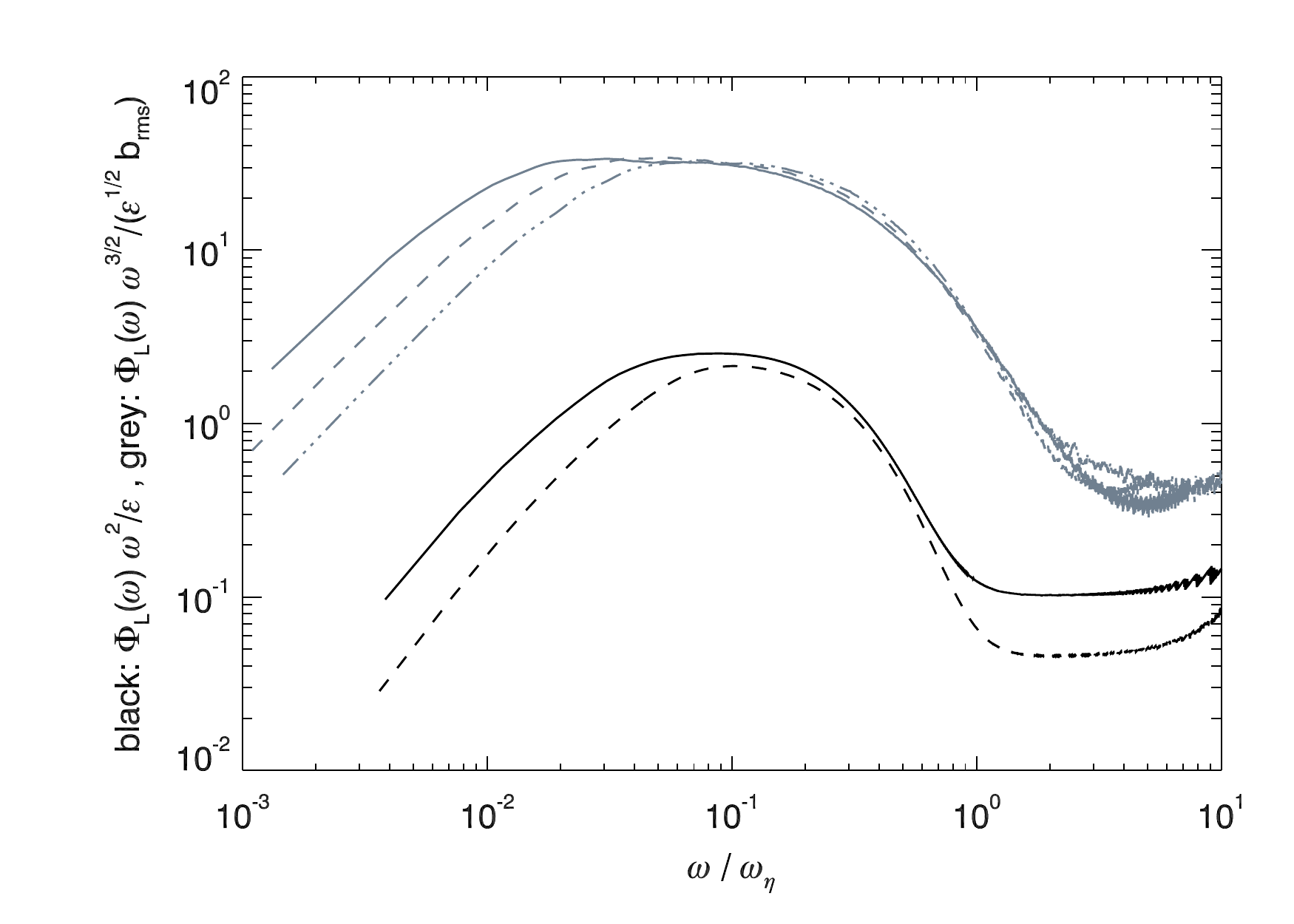}
\caption{\label{spec-b0}{
The compensated Lagrangian frequency spectra in the macroscopically
isotropic three-dimensional (black lines, {compensated by $\omega^{2}/\varepsilon$}) and two-dimensional  (grey lines, {compensated by }$\omega^{3/2}/(\varepsilon^{1/2} b_\text{rms})$)
cases for various Reynolds numbers: $1050$ (black dashed),
$3150$ (black continuous), $2150$ (grey dash-dotted), $6200$ (grey dashed)
and $18240$ (grey continuous). The frequency axis is normalized by the
Kolmogorov frequency $\omega_{\eta}=\pi \sqrt{\epsilon^{K}/\nu}$.
For clarity the spectra for the two-dimensional cases have been shifted
by a constant factor.}}
\end{figure}
In the macroscopically isotropic 3D MHD case the Lagrangian frequency spectrum also
shows a scaling with $\omega^{-2}$ (see Fig.\ \ref{spec-b0}). This supports the GS phenomenology
with its strong cascade
mechanism for this configuration.
In contrast,
the Lagrangian spectra
from the two-dimensional simulations which are shown in the same figure
display an approximate $\omega^{-3/2}$-scaling indicative of a weak interaction cascade. This observation
corroborates the IK phenomenology for the two-dimensional configuration.  
\begin{figure}
\includegraphics[width=8cm]{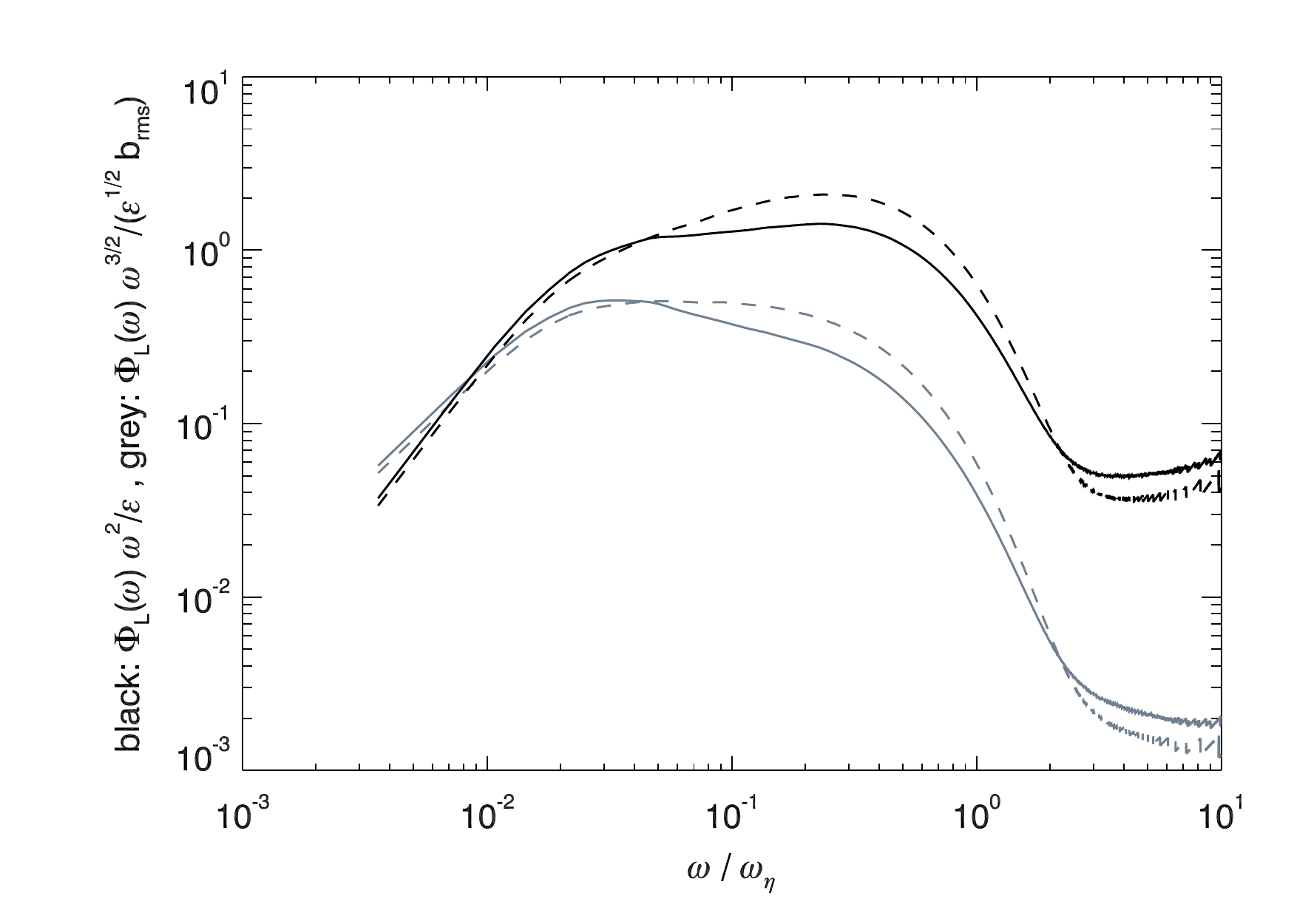}
\caption{\label{spec-b5}The compensated Lagrangian frequency spectrum {(cf. caption of Fig. \ref{spec-b0})} in the MHD case with a strong
mean magnetic field for Reynolds numbers of about $4410$ (cf. table \ref{param-table}). 
Continuous lines: $\vec{b}_{0}$-perpendicular components; dashed lines $\vec{b}_{0}$-parallel component.}
\end{figure}
In the anisotropic MHD case a dependence of the scaling 
on the component of
the fluctuations is observed (see Fig.\ \ref{spec-b5}). For the  
component  along $\vec{b}_0$ $\omega^{-3/2}$-scaling is
observed whereas for the perpendicular components a
scaling exponent
 $\alpha=1.8\pm 0.1$  (where $\Phi(\omega)\sim \omega^{-\alpha}$)
is measured. 
This suggests that either the energy cascade changes its dynamical 
character from weak interaction (field-parallel) to strong interaction (field-perpendicular),
or that the energetic structure of the flow can not be captured adequately by 
a simple 
parallel/perpendicular decomposition.

In summary, a new relation is introduced that relates the nonlinear 
auto-correlation time 
and the cascade time of turbulence with the Lagrangian frequency spectrum.
The relation is corroborated by comparing high-Reynolds-number direct numerical simulations of 
two- and three-dimensional MHD turbulence with currently accepted 
phenomenological expectations. 
As the Lagrangian frequency spectrum is sensitive
to the underlying  cascade mechanism it provides additional insight in 
the yet not fully understood case of MHD turbulence in a strong mean magnetic field.
This is particularly useful in cases where the 
discrimination {between} different theoretical models of turbulence is hard or even impossible to
achieve by Eulerian two-point statistics. The ability to investigate basic 
characteristics of turbulent energy transfer adds significant value to this approach.
\section{Acknowledgements}
The work of G.G. was supported
by Georgian NSF grant ST06/4096 and INTAS grant 06-1000017-9258.


\begin{thebibliography}{25}
\expandafter\ifx\csname natexlab\endcsname\relax\def\natexlab#1{#1}\fi
\expandafter\ifx\csname bibnamefont\endcsname\relax
  \def\bibnamefont#1{#1}\fi
\expandafter\ifx\csname bibfnamefont\endcsname\relax
  \def\bibfnamefont#1{#1}\fi
\expandafter\ifx\csname citenamefont\endcsname\relax
  \def\citenamefont#1{#1}\fi
\expandafter\ifx\csname url\endcsname\relax
  \def\url#1{\texttt{#1}}\fi
\expandafter\ifx\csname urlprefix\endcsname\relax\def\urlprefix{URL }\fi
\providecommand{\bibinfo}[2]{#2}
\providecommand{\eprint}[2][]{\url{#2}}

\bibitem[{\citenamefont{Biskamp}(2003)}]{biskamp:book3}
\bibinfo{author}{\bibfnamefont{D.}~\bibnamefont{Biskamp}},
  \emph{\bibinfo{title}{Magnetohydrodynamic Turbulence}}
  (\bibinfo{publisher}{Cambridge University Press},
  \bibinfo{address}{Cambridge}, \bibinfo{year}{2003}).

\bibitem[{\citenamefont{Frisch}(1996)}]{frisch:book}
\bibinfo{author}{\bibfnamefont{U.}~\bibnamefont{Frisch}},
  \emph{\bibinfo{title}{Turbulence}} (\bibinfo{publisher}{Cambridge University
  Press}, \bibinfo{address}{Cambridge}, \bibinfo{year}{1996}).

\bibitem[{\citenamefont{Iroshnikov}(1964)}]{iroshnikov:ikmodel}
\bibinfo{author}{\bibfnamefont{P.~S.} \bibnamefont{Iroshnikov}},
  \bibinfo{journal}{Soviet Astronomy} \textbf{\bibinfo{volume}{7}},
  \bibinfo{pages}{566} (\bibinfo{year}{1964}), \bibinfo{note}{[Astron. Zh.,
  {40}:742, 1963]}.

\bibitem[{\citenamefont{Kraichnan}(1965)}]{kraichnan:ikmodel}
\bibinfo{author}{\bibfnamefont{R.~H.} \bibnamefont{Kraichnan}},
  \bibinfo{journal}{Physics of Fluids} \textbf{\bibinfo{volume}{8}},
  \bibinfo{pages}{1385} (\bibinfo{year}{1965}).

\bibitem[{\citenamefont{Goldreich and Sridhar}(1995)}]{goldreich_sridhar:gs2}
\bibinfo{author}{\bibfnamefont{P.}~\bibnamefont{Goldreich}} \bibnamefont{and}
  \bibinfo{author}{\bibfnamefont{S.}~\bibnamefont{Sridhar}},
  \bibinfo{journal}{Astrophysical Journal} \textbf{\bibinfo{volume}{438}},
  \bibinfo{pages}{763} (\bibinfo{year}{1995}).

\bibitem[{\citenamefont{Boldyrev}(2006)}]{boldyrev:boldymodelII}
\bibinfo{author}{\bibfnamefont{S.}~\bibnamefont{Boldyrev}},
  \bibinfo{journal}{Physical Review Letters} \textbf{\bibinfo{volume}{96}},
  \bibinfo{pages}{115002} (\bibinfo{year}{2006}).

\bibitem[{\citenamefont{Gogoberidze}(2007)}]{gogoberidze:gogomodel}
\bibinfo{author}{\bibfnamefont{G.}~\bibnamefont{Gogoberidze}},
  \bibinfo{journal}{Physics of Plasmas} \textbf{\bibinfo{volume}{14}},
  \bibinfo{pages}{022304} (\bibinfo{year}{2007}).

\bibitem[{\citenamefont{Biskamp{ \nop{g}and H.
  Welter}}(1989)}]{biskamp_welter:2dmhddecay}
\bibinfo{author}{\bibfnamefont{D.}~\bibnamefont{Biskamp{ \nop{g}and H.
  Welter}}}, \bibinfo{journal}{Physics of Fluids B}
  \textbf{\bibinfo{volume}{1}}, \bibinfo{pages}{1964} (\bibinfo{year}{1989}).

\bibitem[{\citenamefont{M{\"u}ller et~al.}(2003)\citenamefont{M{\"u}ller,
  Biskamp, and Grappin}}]{mueller_biskamp_grappin:anisomhd}
\bibinfo{author}{\bibfnamefont{W.-C.} \bibnamefont{M{\"u}ller}},
  \bibinfo{author}{\bibfnamefont{D.}~\bibnamefont{Biskamp}}, \bibnamefont{and}
  \bibinfo{author}{\bibfnamefont{R.}~\bibnamefont{Grappin}},
  \bibinfo{journal}{Physical Review E} \textbf{\bibinfo{volume}{67}},
  \bibinfo{pages}{066302} (\bibinfo{year}{2003}).

\bibitem[{\citenamefont{Maron and Goldreich}(2001)}]{maron_goldreich:anisomhd}
\bibinfo{author}{\bibfnamefont{J.}~\bibnamefont{Maron}} \bibnamefont{and}
  \bibinfo{author}{\bibfnamefont{P.}~\bibnamefont{Goldreich}},
  \bibinfo{journal}{Astrophysical Journal} \textbf{\bibinfo{volume}{554}},
  \bibinfo{pages}{1175} (\bibinfo{year}{2001}).

\bibitem[{\citenamefont{M{\"u}ller and Grappin}(2005)}]{mueller_grappin:endyn}
\bibinfo{author}{\bibfnamefont{W.-C.} \bibnamefont{M{\"u}ller}}
  \bibnamefont{and} \bibinfo{author}{\bibfnamefont{R.}~\bibnamefont{Grappin}},
  \bibinfo{journal}{Physical Review Letters} \textbf{\bibinfo{volume}{95}},
  \bibinfo{pages}{114502} (\bibinfo{year}{2005}).

\bibitem[{\citenamefont{Mason et~al.}(2008)\citenamefont{Mason, Cattaneo, and
  Boldyrev}}]{mason_cattaneo_boldyrev:specmeas512}
\bibinfo{author}{\bibfnamefont{J.}~\bibnamefont{Mason}},
  \bibinfo{author}{\bibfnamefont{F.}~\bibnamefont{Cattaneo}}, \bibnamefont{and}
  \bibinfo{author}{\bibfnamefont{S.}~\bibnamefont{Boldyrev}},
  \bibinfo{journal}{Physical Review E} \textbf{\bibinfo{volume}{77}},
  \bibinfo{pages}{036403} (\bibinfo{year}{2008}).

\bibitem[{\citenamefont{Perez and Boldyrev}(2009)}]{perez_boldyrev:crosshelmhd}
\bibinfo{author}{\bibfnamefont{J.~C.} \bibnamefont{Perez}} \bibnamefont{and}
  \bibinfo{author}{\bibfnamefont{S.}~\bibnamefont{Boldyrev}},
  \bibinfo{journal}{Physical Review Letters} \textbf{\bibinfo{volume}{102}},
  \bibinfo{pages}{025003} (\bibinfo{year}{2009}).

\bibitem[{\citenamefont{Kaneda}(1993)}]{kaneda:eullag}
\bibinfo{author}{\bibfnamefont{Y.}~\bibnamefont{Kaneda}},
  \bibinfo{journal}{Physics of Fluids {A}} \textbf{\bibinfo{volume}{5}},
  \bibinfo{pages}{2835} (\bibinfo{year}{1993}).

\bibitem[{\citenamefont{Leslie}(1983)}]{leslie:book}
\bibinfo{author}{\bibfnamefont{D.~C.} \bibnamefont{Leslie}},
  \emph{\bibinfo{title}{Developments in the Theory of Turbulence}}
  (\bibinfo{publisher}{Clarendon Press}, \bibinfo{address}{Oxford},
  \bibinfo{year}{1983}).

\bibitem[{\citenamefont{Kraichnan}(1959)}]{kraichnan:dia}
\bibinfo{author}{\bibfnamefont{R.~H.} \bibnamefont{Kraichnan}},
  \bibinfo{journal}{Journal of Fluid Mechanics} \textbf{\bibinfo{volume}{5}},
  \bibinfo{pages}{497} (\bibinfo{year}{1959}).

\bibitem[{\citenamefont{Edwards}(1964)}]{edwards:generalrandomphase}
\bibinfo{author}{\bibfnamefont{S.~F.} \bibnamefont{Edwards}},
  \bibinfo{journal}{Journal of Fluid Mechanics} \textbf{\bibinfo{volume}{18}},
  \bibinfo{pages}{239} (\bibinfo{year}{1964}).

\bibitem[{\citenamefont{Tennekes{ and J. L.
  Lumley}}(1972)}]{tennekes:freqspec2}
\bibinfo{author}{\bibfnamefont{H.}~\bibnamefont{Tennekes{ and J. L. Lumley}}},
  \emph{\bibinfo{title}{A {First} {C}ourse in {Turbulence}}}
  (\bibinfo{publisher}{MIT Press}, \bibinfo{address}{Cambridge, Massachusetts},
  \bibinfo{year}{1972}), \bibinfo{note}{p. 277}.

\bibitem[{\citenamefont{Tennekes}(1975)}]{tennekes:freqspec}
\bibinfo{author}{\bibfnamefont{H.}~\bibnamefont{Tennekes}},
  \bibinfo{journal}{Journal of Fluid Mechanics} \textbf{\bibinfo{volume}{67}},
  \bibinfo{pages}{561} (\bibinfo{year}{1975}).

\bibitem[{\citenamefont{Yeung et~al.}(2006)\citenamefont{Yeung, Pope, and
  Sawford}}]{yeung_pope_sawford:2048simul}
\bibinfo{author}{\bibfnamefont{P.~K.} \bibnamefont{Yeung}},
  \bibinfo{author}{\bibfnamefont{S.~B.} \bibnamefont{Pope}}, \bibnamefont{and}
  \bibinfo{author}{\bibfnamefont{B.~L.} \bibnamefont{Sawford}},
  \bibinfo{journal}{Journal of Turbulence} \textbf{\bibinfo{volume}{7}},
  \bibinfo{pages}{1} (\bibinfo{year}{2006}).

\bibitem[{\citenamefont{Busse et~al.}(2007)\citenamefont{Busse, M{\"u}ller,
  Homann, and Grauer}}]{busse_mueller_homann_grauer:lagradisp}
\bibinfo{author}{\bibfnamefont{A.}~\bibnamefont{Busse}},
  \bibinfo{author}{\bibfnamefont{W.-C.} \bibnamefont{M{\"u}ller}},
  \bibinfo{author}{\bibfnamefont{H.}~\bibnamefont{Homann}}, \bibnamefont{and}
  \bibinfo{author}{\bibfnamefont{R.}~\bibnamefont{Grauer}},
  \bibinfo{journal}{Physics of Plasmas} \textbf{\bibinfo{volume}{14}},
  \bibinfo{pages}{122303} (\bibinfo{year}{2007}).

\bibitem[{\citenamefont{Uhlenbeck and
  Ornstein}(1930)}]{ornstein_uhlenbeck:process}
\bibinfo{author}{\bibfnamefont{G.~E.} \bibnamefont{Uhlenbeck}}
  \bibnamefont{and} \bibinfo{author}{\bibfnamefont{L.~S.}
  \bibnamefont{Ornstein}}, \bibinfo{journal}{Physical Review}
  \textbf{\bibinfo{volume}{36}}, \bibinfo{pages}{823} (\bibinfo{year}{1930}).

\bibitem[{\citenamefont{Eswaran and Pope}(1988)}]{eswaran_pope:forcing}
\bibinfo{author}{\bibfnamefont{V.}~\bibnamefont{Eswaran}} \bibnamefont{and}
  \bibinfo{author}{\bibfnamefont{S.~B.} \bibnamefont{Pope}},
  \bibinfo{journal}{Computers \& Fluids} \textbf{\bibinfo{volume}{16}},
  \bibinfo{pages}{257} (\bibinfo{year}{1988}).

\bibitem[{\citenamefont{Yeung and Pope}(1988)}]{yeung_pope:trackalgo}
\bibinfo{author}{\bibfnamefont{P.~K.} \bibnamefont{Yeung}} \bibnamefont{and}
  \bibinfo{author}{\bibfnamefont{S.~B.} \bibnamefont{Pope}},
  \bibinfo{journal}{Journal of Computational Physics}
  \textbf{\bibinfo{volume}{79}}, \bibinfo{pages}{373} (\bibinfo{year}{1988}).

\bibitem[{\citenamefont{Mordant et~al.}(2004)\citenamefont{Mordant,
  L\'ev\^eque, and Pinton}}]{mordant_leveque_pinton:experiment}
\bibinfo{author}{\bibfnamefont{N.}~\bibnamefont{Mordant}},
  \bibinfo{author}{\bibfnamefont{E.}~\bibnamefont{L\'ev\^eque}},
  \bibnamefont{and} \bibinfo{author}{\bibfnamefont{J.-F.}
  \bibnamefont{Pinton}}, \bibinfo{journal}{New Journal of Physics}
  \textbf{\bibinfo{volume}{6}}, \bibinfo{pages}{1} (\bibinfo{year}{2004}).

\end{thebibliography}
\newcommand{\nop}[1]{}

\end{document}